\documentclass[fleqn,10pt]{wlscirep}
\usepackage[utf8]{inputenc}
\usepackage[T1]{fontenc}
\usepackage{xcolor}
\usepackage{makecell}

\title{PALM: Open Fundus Photograph Dataset with Pathologic Myopia Recognition and Anatomical Structure Annotation}

\author[2,$\dag$]{Huihui Fang}
\author[1,$\dag$]{Fei Li}
\author[2]{Junde Wu}
\author[3]{Huazhu Fu}
\author[2]{Xu Sun}
\author[4]{Jos\'e Ignacio Orlando}
\author[5]{Hrvoje Bogunovi\'c}
\author[1,*]{Xiulan Zhang}
\author[2,*]{Yanwu Xu}

\affil[1]{ State Key Laboratory of Ophthalmology, Zhongshan Ophthalmic Center, Sun Yat-sen University, Guangdong Provincial Key Laboratory of Ophthalmology and Visual Science,Guangzhou, China}
\affil[2]{Intelligent Healthcare Unit, Baidu Inc., Beijing, China}
\affil[3]{Institute of High Performance Computing, Agency for Science, Technology and Research, Singapore}
\affil[4]{Yatiris Group, PLADEMA Institute, CONICET, UNICEN, Tandil, Argentina}
\affil[5]{Christian Doppler Lab for Artificial Intelligence in Retina, Department of Ophthalmology and Optometry, Medical University of Vienna, Vienna, Austria}

\affil[*]{corresponding authors: Yanwu Xu (xuyanwu@baidu.com); Xiulan Zhang (zhangxl2@mail.sysu.edu.cn)} 

\affil[$\dag$]{these authors contributed equally to this work}

\begin{abstract}
Pathologic myopia (PM) is a common blinding retinal degeneration suffered by highly myopic population. Early screening of this condition can reduce the damage caused by the associated fundus lesions and therefore prevent vision loss. Automated diagnostic tools based on artificial intelligence methods can benefit this process by aiding clinicians to identify disease signs or to screen mass populations using color fundus photographs as inputs. This paper provides insights about PALM, our open fundus imaging dataset for pathological myopia recognition and anatomical structure annotation. Our databases comprises 1200 images with associated labels for the pathologic myopia category and manual annotations of the optic disc, the position of the fovea and delineations of lesions such as patchy retinal atrophy (including peripapillary atrophy) and retinal detachment. In addition, this paper elaborates on other details such as the labeling process used to construct the database, the quality and characteristics of the samples and provides other relevant usage notes.
\end{abstract}
\begin{document}

\flushbottom
\maketitle

\thispagestyle{empty}


\section*{Background \& Summary}
\begin{figure}[h!]
\centering
\includegraphics[width=0.75\linewidth]{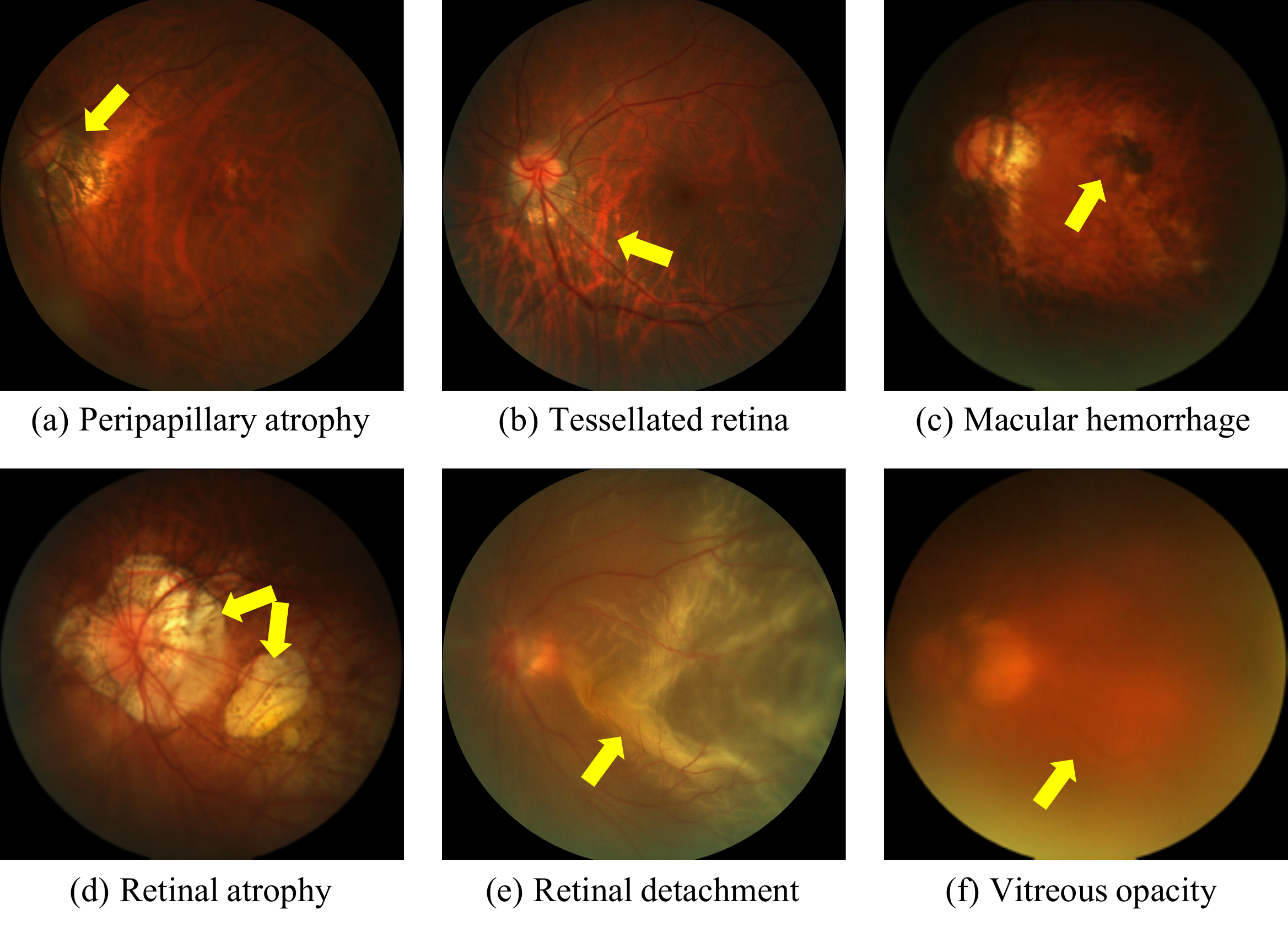}
\caption{Examples of retinal lesions commonly observed in PM cases: (a) Peripapillary atrophies, which occur at the proximity of the optic disc; (b) Tessellated retina, with an observable large choroidal vessels at the posterior fundus pole; (c) Macular hemorrhages, mostly along the crack itself and near from the center of the fovea or in its immediate vicinity; (d) Retinal atrophy, pigment clumping in and around the lesion due to migration of the degenerated retinal pigment epithelium cells into the inner retinal layers; (e) Retinal detachment, an emergency situation in which the retina is pulled away from its normal position; (f) Vitreous opacity, in which the vitreous shrinks and forms strands that cast shadows on the retina. All images corresponds to training samples from PALM.}
\label{fig:lesions}
\end{figure}

Myopia has become a global burden of public health. In 2020, this condition affected nearly 30\% of the world population, and that number is expected to rise up to 50\% by 2050~\cite{sankaridurg2021imi}. Among myopic patients, about 10\% have high myopia~\cite{sankaridurg2021imi}, which is defined by a refractive error of at least -6.00D or an axial length of 26.5 mm or larger~\cite{percival1987redefinition}. As myopic refraction increases, there is an associated risk of pathological changes to the retina and choroid, i.e., high myopia will develop into pathological myopia (PM)~\cite{ohno2021imi}. PM is characterized by the formation of pathologic changes at the posterior pole and the optic disc and by myopic maculopathy~\cite{vingolo2019pathologic}. Among lesions usually observed in PM retinas, some of the more commonly seen are peripapillary atrophies, tessellations and macular hemorrhages (Figure.~\ref{fig:lesions}). These abnormalities can be observed in color fundus photography (CFP) (Figure~\ref{fig:lesions}), which is currently the most cost-effective imaging modality for this condition~\cite{orlando2020refuge}. As undetected PM might potentially result in irreversible visual impairment, it turns relevant to diagnose it at an early stage, to ensure regular patient follow-ups and treatments before further complications. 

In view of the recent developments in artificial intelligence (AI) technology in the field of computer-aided disease diagnosis and treatment, multiple studies started to focus on applying this technology for automated analysis of CFPs~\cite{li2021applications,hagiwara2018computer,sengupta2020ophthalmic}. However, only a few studies aimed at PM in particular. We believe this is likely due to the fact that these data-driven models need to be trained using large curated and annotated datasets, which are currently scarce and not publicly available for this specific condition.

To facilitate future research in this topic, we provide PALM, an open database containing 1200 color fundus photographs related to PM~\cite{fu2019palm}. Unlike other disease datasets already available for the ophthalmic image analysis community such as SCES~\cite{baskaran2015prevalence}, ODIR~\cite{ODIR-5K} or AIROGS~\cite{de_vente_coen_2021_5745834}, ours includes not only CFPs and the disease labels but also optic disc segmentations, the location of the fovea and manual delineations of disease related lesions. These additional  annotations can assist in building complementary AI models for disease classification and interpretation, which can aid clinicians to comprehensively analyze disease patterns and provide a more accurate diagnostic of PM. 


PALM dataset has been released as part of the PAthologicaL Myopia challenge, which was held in conjunction with the International Symposium on Biomedical Imaging (ISBI) in 2019. To date, our dataset has already been used in more than 100 papers in the field of automated diagnosis of PM~\cite{biswas2022color,app11020591,cui2021pathological,rauf2021automatic,hemelings2021pathological} or fundus structure analysis~\cite{guo2020lesion,9193942,du2021deep} based on CFPs.

\section*{Methods}
\subsection*{Data Collection}

\begin{figure}[h!]
    \centering
    \includegraphics[width=0.75\linewidth]{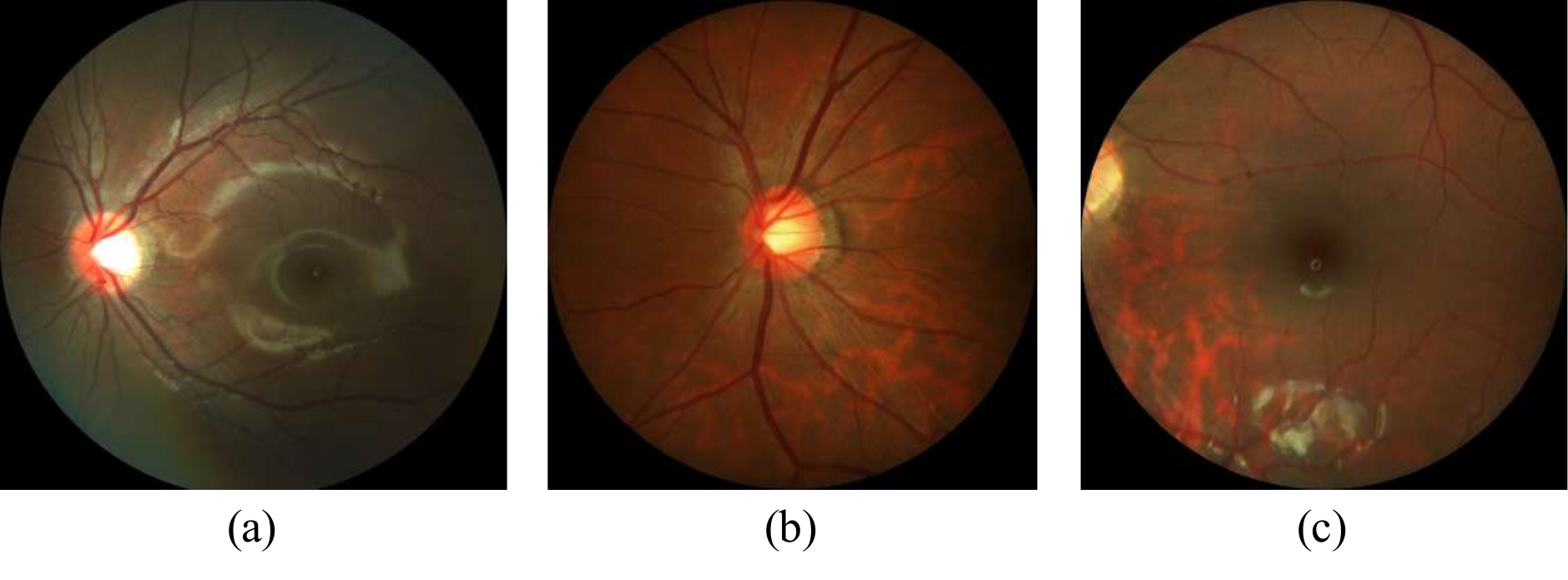}
    \caption{Color fundus images from PALM training set centered at (a) the midpoint between optic disc and fovea, (b) the optic disc, and (c) the fovea).}
    \label{fig:field_of_view}
\end{figure}

PALM contains retinal images retrospectively collected from a myopic examination cohort at the Zhongshan Ophthalmic Center (ZOC), Sun Yat-sen University, China. Each CFP was acquired in a single field of view, i.e., the fundus was photographed with the midpoint of the optic disc (OD) and macula as the center, or in a dual field of view, i.e. with the OD and macula as the center of the image, respectively (Figure.~\ref{fig:field_of_view}). 

The protocol for retrieving the images was approved by the ethics board of ZOC. CFPs were included if: (1) they were acquired with a single field of view or with a dual field of view; (2) they show no image quality problems or minor exposure issues (minor smudges, out-of-focus or blurring that do not affect overall image interpretation were tolerated). 
Images were excluded if they showed any trace of treatment, severe exposure abnormalities, severe refractive interstitial opacities, large-scale contaminations or if information about its origin was missing.


The final dataset comprises 1200 fundus images from 720 subjects (Male: 48.1\%, mean age: 37.5 $\pm$ 15.91), captured with either a Zeiss Visucam 500 camera (1047 images, with a resolution of $2124 \times 2056$ pixels) or a Canon CR-2 camera (153 images, with a resolution of $1444 \times 1444$ pixels). The database is provided already split in a training, a validation and a test set (Table~\ref{tab:data_intro}), with images belonging to the same patient assigned to the same set.

\subsection*{Disease Diagnosis}

\begin{figure}[h!]
    \centering
    \includegraphics[width=\linewidth]{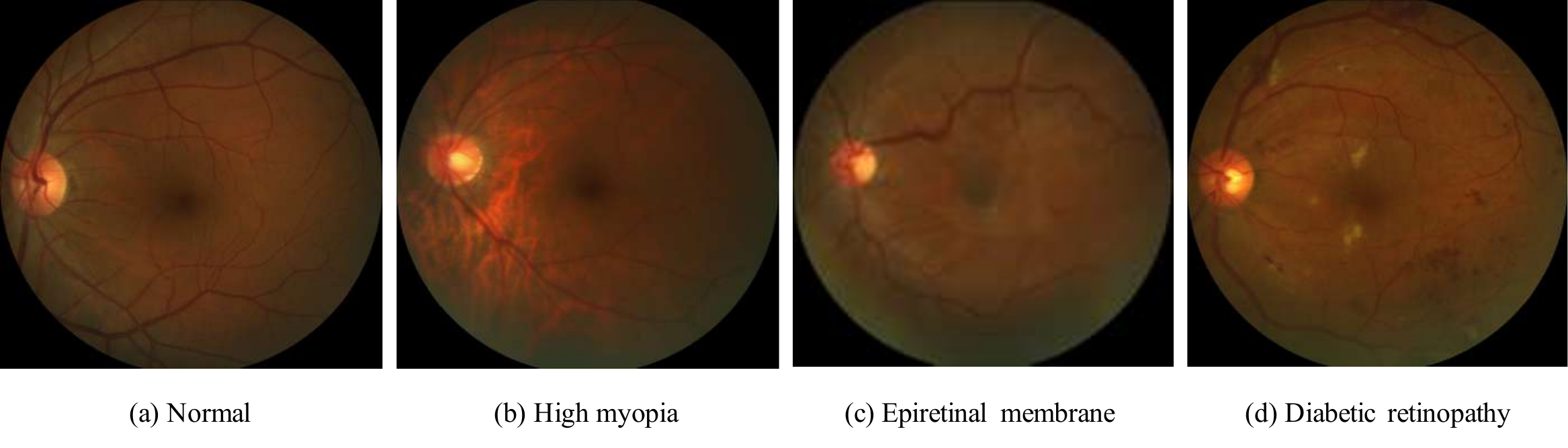}
    \caption{Examples of color fundus images from PALM corresponding to the non-PM category. Notice that this subset contains not only healthy individuals (a) but also subjects with other conditions such as high myopia (b), epiretinal membrane (c) and diabetic retinopathy (d), among others.}
    \label{fig:non-pm}
\end{figure}

Disease labels indicating presence or absence of PM were assigned to each scan based on its associated clinical records, which considered in a comprehensive manner the medical history, refractive error, fundus imaging reports, etc. The guidelines of the International Myopia Institute~\cite{flitcroft2019imi} were followed, so that an individual was considered as PM if structural changes in the posterior segment of the eye caused by an excessive axial elongation associated with myopia were observed, including posterior staphyloma, myopic maculopathy, and high myopia-associated glaucoma-like optic neuropathy. These alterations were observed during clinical examination using multiple imaging modalities, including optical coherence tomography (OCT), fluorescein angiography (FA), and OCT angiography (OCTA). Notice that non-PM images might not necessarily correspond to healthy individuals, as shown in Figure.~\ref{fig:non-pm}.


\subsection*{Manual Annotations}

\begin{figure}[h!]
    \centering
    \includegraphics[width=\linewidth]{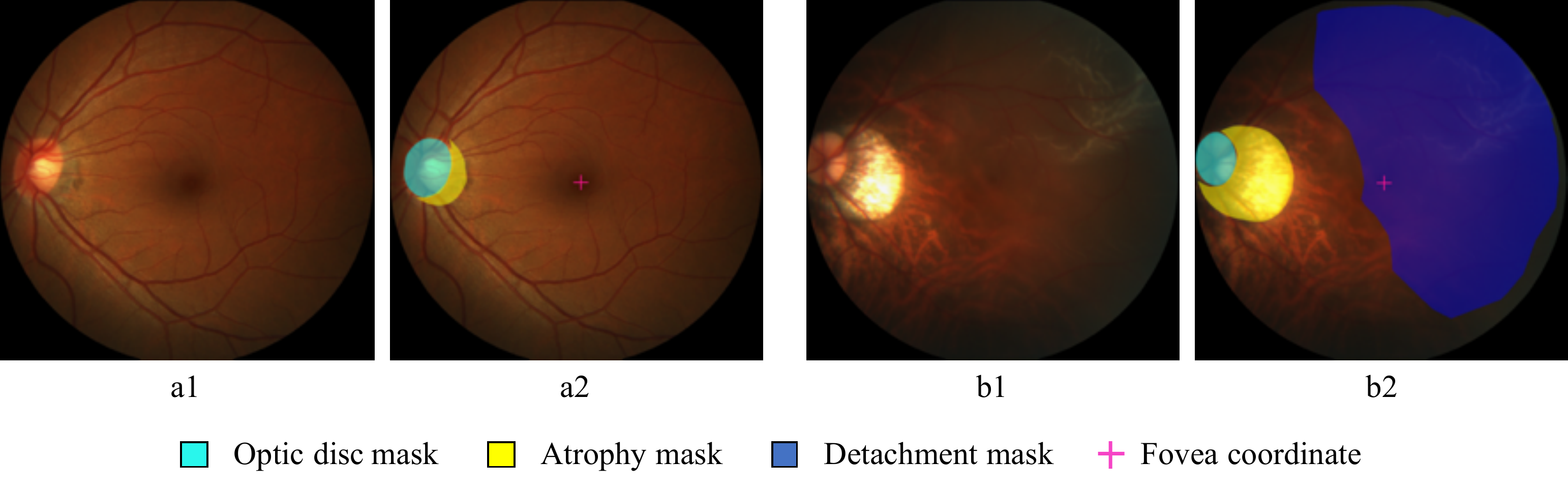}
    \caption{Example of the annotation interface used by the experts in (a) a no-PM sample, and (b) a PM sample. (a1) and (b1): original input images, (a2) and (b2): manual annotations.}
    \label{fig:annotation}
\end{figure}

\begin{figure}[h!]
\centering
\includegraphics[width=0.8\linewidth]{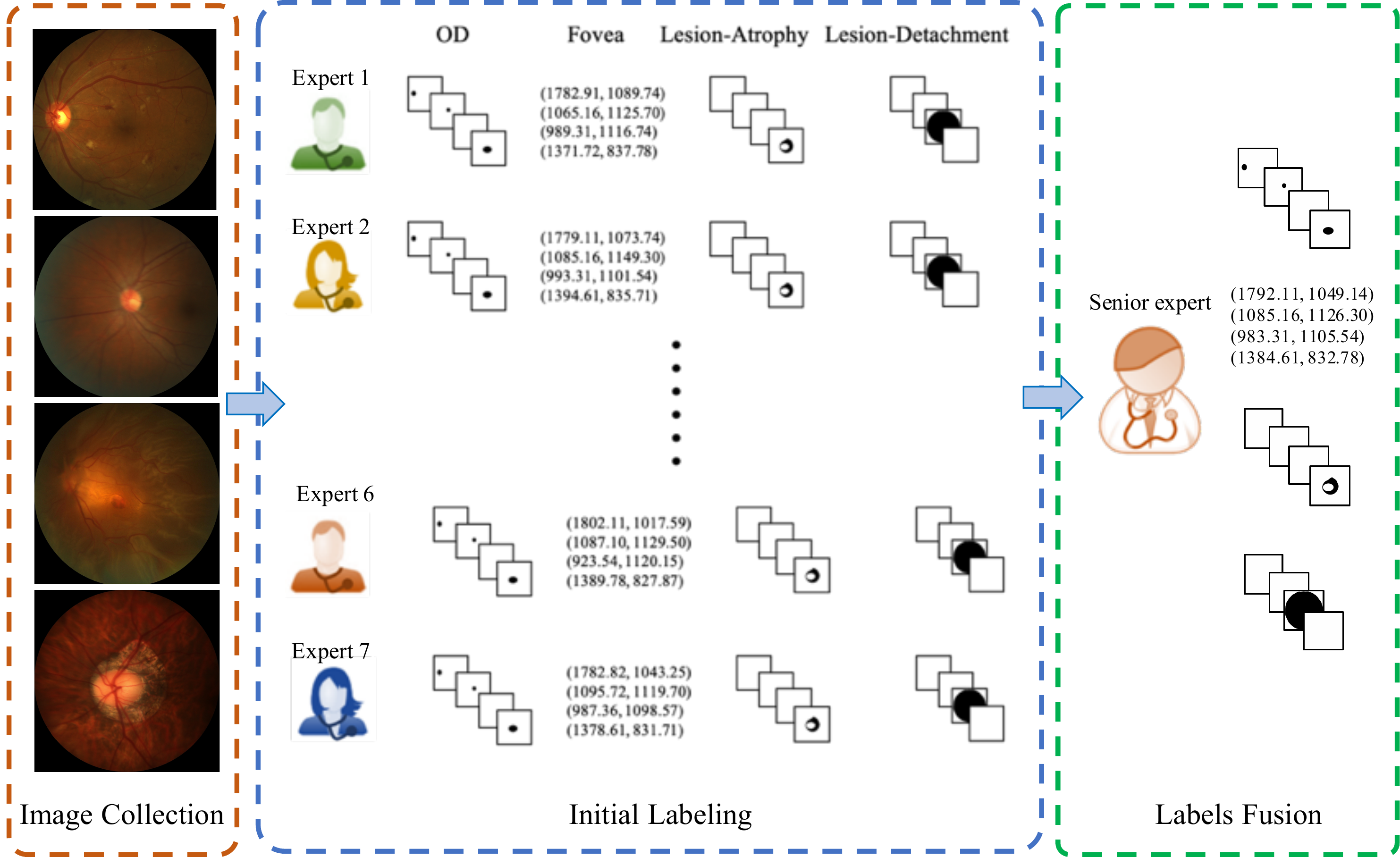}
\caption{Manual annotation process. Manual delineations were performed by seven different experts and reviewed subsequently by one senior expert. 
}
\label{fig:data_collection_labelling}
\end{figure}

Manual delineations of the optic disc and fundus lesions and the annotation of the fovea localization (Figure.~\ref{fig:annotation}) were performed by seven ophthalmologists with an average experience of 8 years in the field (ranging from 5 to 10 years) and one senior ophthalmologist, with more than 10 years of experience, all of them part of ZOC staff (Figure.~\ref{fig:data_collection_labelling}). All ophthalmologists annotated the structures by themselves without having access to any patient information or knowledge of disease prevalence in the data. Details regarding the annotation protocol followed for each specific target are provided in the sequel.

\subsubsection*{Optic disc annotation}

Experts used a free annotation tool with capabilities for image review, zoom, contrast enhancement, and circle and ellipse fitting, to manually draw elliptical structures approximately covering the optic disc. Pixels within the fitted area were then mapped to a binary pixel-wise segmentation mask. Annotations of the same image performed by the seven different graders were merged into a single one by majority voting. The senior ophthalmologist then performed a quality check of this resulting mask to account for any potential mistakes. When errors in the annotations were observed, the senior ophthalmologist analyzed each of the seven masks, removed those that were considered erroneous and repeated the majority voting process with the remaining ones. 

\subsubsection*{Fovea localization}

A tool that allows to manually set the position of the crosshair on an image was used to approximate the location of the fovea. The final annotation was produced by averaging the seven coordinates provided by the ophthalmologists, which was further reviewed by the senior ophthalmologist. In case certain fovea coordinates in the set were considered erroneous, this expert removed them and repeated the averaging process with the remaining ones.

\subsubsection*{Lesion annotation}

Two types of lesions related to PM were annotated on each image: patchy retinal atrophy (including peripapillary atrophy) and retinal detachment. Experts used the same annotation tool as for the optic disc, although using a closed curve to outline the lesions. Unlike the SUSTech-SYSU dataset~\cite{lin2020sustech}, a free-form closed curve was used to allow a more accurate approximation of lesion borders. The same revision process used for the optic disc mask was followed for lesion masks as well.

\subsection*{Data Validation}

Data quality was automatically verified using the Multiple Color-space Fusion Network (MCF-Net) approach by Fu \textit{et al.}~\cite{fu2019evaluation}, which classifies color fundus images into quality grades good, usable and reject based on different color-space representation at feature and prediction levels. Tables~\ref{tab:fu-subclass} and~\ref{tab:fu-subsets} indicate the number of images grouped by quality according to each disease label and for each split, respectively.

According to Table~\ref{tab:fu-subclass}, 93.8\% of the images in the non-PM category are classified either as good or usable, while this number reduces to 5.8\% in the PM category. This is related to the characteristics of the PM images. As shown in Figure~\ref{fig:lesions}, retinal detachment, peripapillary atrophy, retinal atrophy and other lesions are common in the fundus images of patients with PM, which interfere with the recognition of anatomical structures such as the optic disc and the macula. Fu et.al.~\cite{fu2019evaluation} model associated images with invisible disc or macula to the ‘Reject’ grade, which is common in other similar studies~\cite{raj2019fundus,shao2017automated,fleming2006automated,niemeijer2006image}. While this explains why the number of "reject" images is observed in the non-PM class, this renders erroneous in a real-world setting, in which images like these should be processed by the subsequent AI models to produce adequate diagnostics. The significance of the propose PALM dataset is therefore increased, as it provides a series of scans that reproduces clinical use-cases currently ignored or neglected by the existing literature.


In addition, it can be seen from Table~\ref{tab:fu-subsets} that the image quality distribution in training, validation and testing subsets of the proposed PALM dataset is relatively consistent. This ensures that the posterior evaluation of the model will not suffer from unstratified sampling and distribution biases.

\section*{Data Records}

PALM is uploaded to FigShare as a compressed file and will be made publicly available by the time of publication. All personal information that could be used to identify the patients was removed before preparation. Data is provided already partitioned in folders \textit{Training set}, \textit{Validation set} and \textit{Test set}, with each subset containing folders for '\textit{Images}', '\textit{Disc Masks}', and '\textit{Lesion Masks}', and three Excel files (i.e. '\textit{Classification Labels.xlsx}', '\textit{Fovea Localization.xlsx}', and '\textit{Supplementary Information.xlsx}').


The '\textit{Images}' folder within each subset contains 400 color fundus images each, stored in JPEG format, with 8 bits per color channel. Similarly, the '\textit{Disc Masks}' has all binary optic disc masks associated to each fundus picture, as BMP files, also with 8 bits per color channel. On the other hand, the ’\textit{Lesion Masks}’ folder contains two subfolders, corresponding to each lesion type target, namely, '\textit{Atrophy}' and '\textit{Detachment}', with binary annotations for patchy retinal atrophy and retinal detachment, respectively. The file format of the lesion segmentation masks is consistent with those of the optic disc masks.


The '\textit{Classification Labels.xlsx}' file contains the labels for PM classification, with 1 representing PM and 0 no-PM. The '\textit{Fovea Localization.xlsx}' provides the x- and y- coordinates of the fovea. Notice that a coordinate $(0, 0)$ is used when the fovea is not visible in the associated image. Additional information, i.e., the equipments used for image acquisition and the type of photo centering, are provided in the ’\textit{Supplementary Information.xlsx}' file, using in one column 1 to denote Zeiss and 2 to denote Canon devices, and, in a second column, 1, 2, and 3 to indicate optic disc centered, fovea centered, and center at the midpoint of optic disc and fovea, respectively.

\section*{Technical Validation}

Tables~\ref{tab:detachment} and~\ref{tab:atropy} provide the proportion of pixels corresponding to regions with retinal detachment and patchy retinal atrophy, respectively, differentiating by each disease category, acquisition protocol and subset. As expected, no retinal detachment lesions were found in images acquired with the optic disc at the center of the field of view or in images of patients with no PM (Table~\ref{tab:detachment}). Patchy retinal atrophies, on the other hand, are observed in both PM and non-PM categories (Table~\ref{tab:atropy}), although their size is much larger in PM subjects. Furthermore, these lesions are more frequently observed in images with visible optic disc, which is expected considering that these lesions appear at the vecinity of this anatomical structure.


In addition to discussing the characteristics of the lesions in the images, we also counted the properties of the fovea position in the fundus images with respect to photo centering used. Table~\ref{tab:fovea_loc} shows the mean $[\bar{x}, \bar{y}]$ of the normalized coordinates for the fovea localization among the fundus images with different photo centering in PALM dataset. $\bar{x} = \frac{1}{N} \sum_{i=1}^{N}\frac{x_i}{W_i}$, $\bar{y} = \frac{1}{N} \sum_{i=1}^{N}\frac{y_i}{H_i}$, where $[x_i, y_i]$ is the coordinate of the fovea in the $i$th image, and $H_i$ and $W_i$ are the height and width of the image. $N$ is the total number of the samples in the corresponding categories. From the table, we can see that in fundus image centered on the optic disc, the fovea appears on the right side of the images, as fundus pictures correspond in all cases to left eyes. In the images centered on the macula, the fovea appears in the center, and in the images centered on the midpoint of the optic disc and the macula, the fovea appears to the right of the image center. Thus, the fovea position characteristics are consistent with our expectations.

\section*{Usage Notes}

Our dataset will be accessible through a FigShare link by the time of publication. 
PALM images can be used to perform studies on automated PM classification, optic disc segmentation, fovea localization, and atrophic lesion retinal detachment segmentation. In the aforementioned PALM Challenge, these tasks were set up as sub-challenges in which different participating teams proposed their own methods to automate them. The evaluation of their corresponding approaches in the validation and test sets for each of the sub-challenges are accessible in \url{https://palm.grand-challenge.org/SemifinalLeaderboard/} and \url{https://palm.grand-challenge.org/Test/}.

For the studies on classification, segmentation and localization, we designed a series of baseline models~\cite{fang2022dataset}, which we trained and evaluated using PALM data. For optic disc and lesion segmentation, we used a standard U-shaped network~\cite{ronneberger2015u} with residual blocks, while for PM classication and fovea localization we utilized ResNet50~\cite{he2016deep} architectures. The corresponding code has been released as open source (See Code availability section). 

Table~\ref{tab:baseline_results} shows the evaluation results of the baseline models on each task. The chosen evaluation metrics are commonly used for classification, segmentation and localization tasks, and are equal to those used in other related ophthalmic image analysis challenges~\cite{orlando2020refuge, fang2022adam}. The baseline classification model achieved good results on both validation and test sets, as observed in Table~\ref{tab:baseline_results}. This is mainly because the fundus images of PM category in our dataset correspond to severe PM, so the associated fundus lesion features were particularly evident. For optic disc segmentation, results are slightly lower than those observed in cases with other conditions such as glaucoma~\cite{orlando2020refuge}. This is likely due to the PM cases showing significant abnormalities within the optic disc region. Furthermore, results for lesion segmentation achieved by the baseline model are poor. This might be a consequence of both the irregular shape of patchy retinal atrophy and retinal detachment lesions (which vary significantly from one case to another) and due to the scarcity of sufficient training samples. In the future, we will aim at collecting more PM images of various processes, and supplement the dataset with additional samples of patchy retinal atrophy and retinal detachment lesions to reduce the impact of this observation.

Tables~\ref{tab:categories_PM_class} to \ref{tab:categories_detach} show the performance of the baseline models stratified by category. As stated above, the baseline model was able to segment the optic disc region more accurately in non-PM samples than in PM cases (Table~\ref{tab:categories_OD_seg}). This is because PM lesions interfere with the overall appearance of the optic disc features, thus leading to poor segmentation prediction results. Therefore, new models should incorporate mechanisms to deal with these characteristics. A similar conclusion can be drawn for the fovea localization task (Table~\ref{tab:categories_fovea}), in which much larger errors are observed in PM cases. This further emphasize the importance of PALM, which complements other existing databases with much more challenging scenarios to further improve existing models. Results for patchy retinal atrophy segmentation (Table~\ref{tab:categories_atrophy}) are much better in PM cases than in non-PM, which can be due to the much more subtle appearance of these lesions in non-PM. Finally, notice that performance in the validation and test sets are similar (Tables~\ref{tab:categories_PM_class}-\ref{tab:categories_detach}), which indicates that our data partition is well-balanced, and suitable for researchers to directly use to tune and test the models. 


In summary, PALM is the first dataset for assisting AI researchers in training AI models for automated PM analysis. Disease labels are complemented by a series of manual annotations of lesions and anatomical structures that can allow studies focusing on exploiting complementary features to enhance results. Furthermore, PALM can be used in combination with other existing fundus image datasets such as REFUGE~\cite{orlando2020refuge} and ADAM~\cite{fang2022adam} to produce much more robust models for optic disc segmentation, fovea localization and even quality assessment.


\section*{Code availability}
The source code for the image quality assessment by \textit{Fu et.al.} can be accessed at \url{https://github.com/hzfu/EyeQ}. The source code for the baseline model training and testing is available at \url{https://github.com/tianyizheming/ichallenge_baseline}.

\bibliography{sample}

\section*{Author contributions statement}

Acquisition of data: X.Z., and F.L.\\
Analysis and interpretation of data: H.F. (Huihui Fang), J.W., and X.S.\\
PALM Challenge organization: Y.X., X.Z., H.F. (Huazhu Fu), F.L., X.S., J.I.O, H.B.\\
Drafting the work or revising it critically: H.F. (Huihui Fang), F.L., J.I.O, H.B., H.F. (Huazhu Fu), J.W., and Y.X.

\section*{Competing interests} 

The authors declare no competing interests.

\section*{Figures \& Tables}

\begin{figure}[h!]
    \centering
    \includegraphics[width=0.35\linewidth]{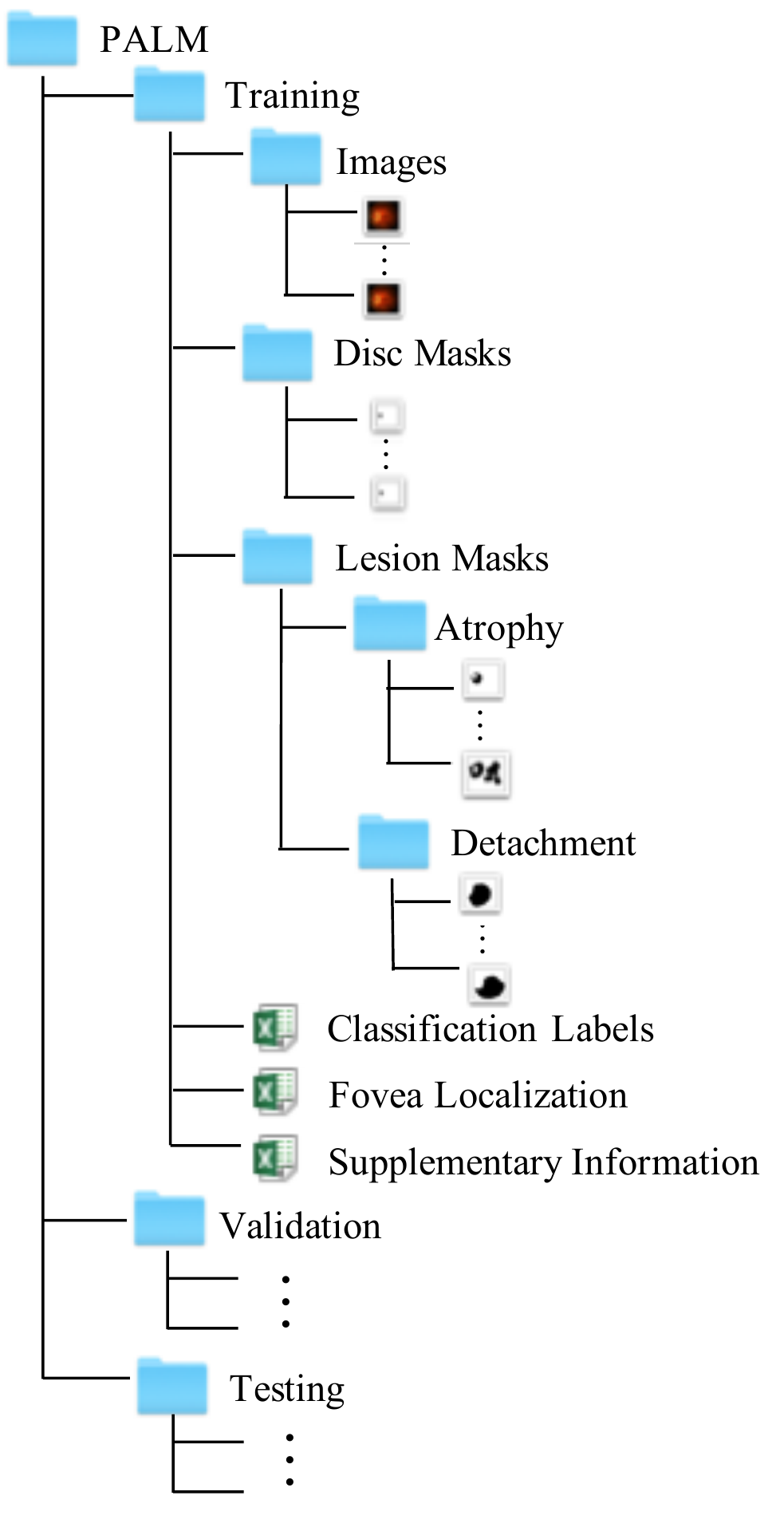}
    \caption{Folder organization of our PALM dataset.}
    \label{fig:data_store}
\end{figure}

\begin{table}[h!]
  \centering
  \caption{Summary of the main characteristics of each subset of the PALM dataset, stratified by disease, structure, lesion, image acquisition type, and acquisition device.}
    \begin{tabular}{ccccccccc}
    \hline
    Set   & Num.  & PM/Non-PM & With/o OD & With/o Fovea & \makecell[c]{With/o \\Detachment} & \makecell[c]{With/o \\ Atrophy} & \makecell[c]{Photo centering \\(OD/ fovea/ \\ midpoint of \\ OD and fovea)} & \makecell[c]{Device \\(Zeiss/Canon)} \\
    \hline
    Training & 400   & 213/187 & 381/19 & 397/3 & 12/388 & 311/89 & 42/258/100      & 350/50 \\
    Validation & 400   & 211/189 & 379/21 & 397/3 & 6/394 & 271/129 &  43/258/99     & 344/56 \\
    Testing & 400   & 213/187 & 384/16 & 398/2 & 6/394 & 288/112 &  38/284/78     & 353/47 \\
    \hline
    Total & 1200  & 637/563 & 1144/56 & 1192/8 & 24/1176 & 870/330 &  123/800/277     & 1047/153 \\
    \hline
    \end{tabular}%
  \label{tab:data_intro}%
\end{table}%


\begin{table}[h!]
\begin{minipage}[c]{0.5\textwidth}
\centering
\captionof{table}{The image quality assessment results in PM and Non-PM samples of the PALM dataset according to the fundus image quality assessment method proposed by Fu et al.~\cite{fu2019evaluation}}
\label{tab:fu-subclass}
\begin{tabular}{cccc}
    \hline
    Quality & Good  & Usable & Reject \\
    \hline
    PM    & 37    & 123   & 477 \\
    Non-PM  & 238   & 290   & 35 \\
    \hline
\end{tabular}
\end{minipage}
\begin{minipage}[c]{0.5\textwidth}
\centering
\captionof{table}{The image quality assessment results in different subsets of the PALM dataset according to the fundus image quality assessment method proposed by Fu et al.~\cite{fu2019evaluation}}
\label{tab:fu-subsets}
\begin{tabular}{cccc}
    \hline
    Quality & Good  & Usable & Reject \\
    \hline
    Training set & 84    & 143   & 173 \\
    Validation set & 87    & 143   & 170 \\
    Testing set & 104   & 127   & 169 \\
    \hline
    \end{tabular}
\end{minipage}
\end{table}


\begin{table}[htbp]
  \centering
  \caption{Proportion of the detachment mask pixels in different categories of fundus images in the PALM dataset.}
    \begin{tabular}{ccccccc}
    \hline
          & \multicolumn{2}{c}{Training} & \multicolumn{2}{c}{Validation} & \multicolumn{2}{c}{Testing} \\
          \cline{2-7}
          & \multicolumn{1}{c}{PM} & \multicolumn{1}{c}{Non-PM} & \multicolumn{1}{c}{PM} & \multicolumn{1}{c}{Non-PM} & \multicolumn{1}{c}{PM} & \multicolumn{1}{c}{Non-PM} \\
          \hline
    Optic disc centered & 0\%     & 0\%     & 0\%     & 0\%     & 0\%     & 0\% \\
    Fovea centered & 3.6\% & 0\%     & 1.2\% & 0\%     & 0.8\% & 0\% \\
    \makecell[c]{Midpoint of the optic disc\\ and fovea centered} & 2.4\% & 0\%     & 2.3\% & 0\%     & 3.6\% & 0\% \\
    \hline
    \end{tabular}%
  \label{tab:detachment}%
\end{table}%

\begin{table}[htbp]
  \centering
  \caption{Proportion of the atrophy mask pixels in different categories of fundus images in the PALM dataset.}
    \begin{tabular}{ccccccc}
    \hline
          & \multicolumn{2}{c}{Training} & \multicolumn{2}{c}{Validation} & \multicolumn{2}{c}{Testing} \\
          \cline{2-7}
          & PM    & Non-PM & PM    & Non-PM & PM    & Non-PM \\
          \hline
    Optic disc centered & 15.1\% & 0.6\% & 16.8\% & 1.1\% & 16.8\% & 1.5\% \\
    Fovea centered & 9.8\% & 0.3\% & 14.3\% & 0.3\% & 13.3\% & 0.3\% \\
    \makecell[c]{Midpoint of the optic disc\\ and fovea centered} & 14.2\% & 0.3\% & 15.7\% & 0.3\% & 20.3\% & 0.5\% \\
    \hline
    \end{tabular}%
  \label{tab:atropy}%
\end{table}%

\begin{table}[htbp]
  \centering
  \caption{The mean $[\bar{x}, \bar{y}]$ of the normalized coordinates for the fovea localization among the fundus images with different photo centering in PALM dataset.}
    \begin{tabular}{cccc}
    \hline
          & Training & Validation & Testing \\
    \hline
    Optic disc centered & [0.856, 0.495] & [0.846, 0.506] & [0.851, 0.515] \\
    Fovea centered & [0.522, 0.513] & [0.524, 0.516] & [0.5224, 0.514] \\
    \makecell[c]{Midpoint of the optic disc \\ and fovea centered} & [0.614, 0.503] & [0.607, 0.503] & [0.607, 0.506] \\
    \hline
    \end{tabular}%
  \label{tab:fovea_loc}%
\end{table}%

\begin{table}[htbp]
  \centering
  \caption{Performance of baseline classification, segmentation and localization models trained using PALM training set, as evaluated in PALM's validation and test sets. OD: Optic Disc. AUC: Area Under the receiver-operating Curve. F1: F1-score. Acc: Accuracy. Sen: Sensitivity. Spe: specificity. ED: Euclidean Distance.}
    \begin{tabular}{cccccccccc}
    \hline
          & \multicolumn{5}{c}{Classification}    & \makecell[c]{OD \\ Segmentation} & \makecell[c]{Fovea \\ Localization} & \makecell[c]{Atrophy \\ Segmentation} & \makecell[c]{Detachment \\Segmentation} \\
          \cline{2-10}
          & AUC   & F1    & Acc   & Sen   & Spe   & DICE  & ED (pixels) & DICE  & DICE\\
          \hline
    Validation  & 0.999 & 0.981 & 0.980  & 0.976 & 0.984 & 0.876$\pm$0.148 & 90.928$\pm$134.942 & 0.675$\pm$0.266 &  0.412$\pm$0.284  \\
    Test  & 0.994 & 0.969 & 0.968 & 0.962 & 0.973 & 0.868$\pm$0.142 & 83.998$\pm$103.3.4 & 0.674$\pm$0.253 & 0.474$\pm$0.272 \\
    \hline
    \end{tabular}%
  \label{tab:baseline_results}%
\end{table}%

\begin{table}[htbp]
  \centering
  \caption{Automated PM classification performance of our baseline trained using PALM and evaluated in the validation and test sets, stratified by disease, acquisition device and image acquisition type, in terms of accuracy (Acc). Acc for PM and Non-PM categories are calculated by $\frac{TP}{N_{PM}}$, and $\frac{TN}{N_{Non-PM}}$, respectively. $N_{PM}$ and $N_{Non-PM}$ are the total numbers of samples in the corresponding categories.} 
    \begin{tabular}{cccccccc}
    \hline
    Acc   & PM    & Non-PM & ZEISS & CANON & Optic disc centered & Fovea centered & \makecell[c]{Midpoint of the optic disc \\ and fovea centered} \\
    \hline
    Validation & 0.976 & 0.984 & 0.98  & 0.982 & 1     & 0.977 & 0.98 \\
    Testing & 0.962 & 0.973 & 0.972 & 0.936 & 0.895 & 0.979 & 0.962 \\
    \hline
    \end{tabular}%
  \label{tab:categories_PM_class}%
\end{table}%

\begin{table}[htbp]
  \centering
  \caption{Automated optic disc segmentation performance of our baseline trained using PALM and evaluated in the validation and test sets, stratified by disease, acquisition device and image acquisition type, in terms of DICE.}
    \begin{tabular}{cccccccc}
    \hline
          & PM    & Non-PM & ZEISS & CANON & \makecell[c]{Optic disc \\ centered} & \makecell[c]{Fovea\\ centered} & \makecell[c]{Midpoint of the\\ optic disc and\\ fovea centered} \\
          \hline
    Validation & 0.831$\pm$0.188 & 0.922$\pm$0.064 & 0.885$\pm$0.137 & 0.802$\pm$0.205 & 0.802$\pm$0.200 & 0.894$\pm$0.122 & 0.865$\pm$0.168 \\
    Testing & 0.820$\pm$0.181  & 0.918$\pm$0.046 & 0.869$\pm$0.145 & 0.861$\pm$0.115 & 0.865$\pm$0.115 & 0.882$\pm$0.123 & 0.821$\pm$0.197 \\
    \hline
    \end{tabular}%
  \label{tab:categories_OD_seg}%
\end{table}%

\begin{table}[htbp]
  \centering
  \caption{Automated fovea localization performance of our baseline trained using PALM and evaluated in the validation and test sets, stratified by disease, acquisition device and image acquisition type, in terms of Euclidean Distance (ED).}
    \begin{tabular}{cccccccc}
    \hline
    ED (pixels) & PM    & Non-PM & ZEISS & CANON & \makecell[c]{Optic disc \\ centered} & \makecell[c]{Fovea\\ centered} & \makecell[c]{Midpoint of the\\ optic disc and\\ fovea centered} \\
    \hline
    Validation & 108.1$\pm$178.3 & 71.8$\pm$48.5 & 80.8$\pm$106.7 & 153.2$\pm$235.8 & 188.3$\pm$339.4 & 71.1$\pm$74.5 & 100.2$\pm$62.0 \\
    Testing & 103.4$\pm$132.8 & 62.0$\pm$42.7 & 77.3$\pm$92.6 & 134.2$\pm$153.6 & 119.6$\pm$62.1 & 76.0$\pm$114.6 & 95.8$\pm$63.5 \\
    \hline
    \end{tabular}%
  \label{tab:categories_fovea}%
\end{table}%

\begin{table}[htbp]
  \centering
  \caption{Automated patchy retinal atrophy segmentation performance of our baseline trained using PALM and evaluated in the validation and test sets, stratified by disease, acquisition device and image acquisition type, in terms of DICE.}
    \begin{tabular}{cccccccc}
    \hline
    DICE  & PM    & Non-PM & ZEISS & CANON & \makecell[c]{Optic disc \\ centered} & \makecell[c]{Fovea\\ centered} & \makecell[c]{Midpoint of the\\ optic disc and\\ fovea centered} \\
    \hline
    Validation & 0.778$\pm$0.182 & 0.358$\pm$0.232 & 0.655$\pm$0.276 & 0.751$\pm$0.209 & 0.796$\pm$0.149 & 0.619$\pm$0.287 & 0.717$\pm$0.243 \\
    Testing & 0.770$\pm$0.171  & 0.407$\pm$0.254 & 0.656$\pm$0.263 & 0.770$\pm$0.163  & 0.772$\pm$0.159 & 0.613$\pm$0.269 & 0.779$\pm$0.193 \\
    \hline
    \end{tabular}%
  \label{tab:categories_atrophy}%
\end{table}%

\begin{table}[htbp]
  \centering
  \caption{Automated retinal detachment segmentation performance of our baseline trained using PALM and evaluated in the validation and test sets, stratified by disease, acquisition device and image acquisition type, in terms of DICE.}
    \begin{tabular}{cccccccc}
    \hline
    DICE  & PM    & Non-PM & ZEISS & CANON & \makecell[c]{Optic disc \\ centered} & \makecell[c]{Fovea\\ centered} & \makecell[c]{Midpoint of the\\ optic disc and\\ fovea centered} \\
    \hline
    Validation & 0.412$\pm$0.284 & -     & 0.412$\pm$0.284 & -     & -     & 0.549$\pm$0.341 & 0.343$\pm$0.221 \\
    Testing & 0.474$\pm$0.272 & -     & 0.474$\pm$0.272 & -     & -     & 0.608$\pm$0.108 & 0.407$\pm$0.302 \\
    
    \hline
    \end{tabular}%
  \label{tab:categories_detach}%
\end{table}%

\end{document}